\documentclass[draft,linenumbers]{agujournal}
\journalname{Geophysical Research Letters}
\usepackage{amsmath}
\usepackage{amssymb}
\usepackage{graphicx}
\usepackage{bm}
\nolinenumbers
\begin{document}
\title{Characterizing Ion Flows Across a Dipolarization Jet}

\authors{H. Arnold \affil{1}, M.~Swisdak \affil{1},
J.~F.~Drake\affil{1}}

\affiliation{1}
{IREAP, University of Maryland, College Park MD 20742-3511, USA}

\begin{keypoints}
\item Dependence of the structure of dipolarization jets, or reconnection exhausts, on the cross-tail width is studied through 2D simulations 
\item For small widths, a linear relationship is found between the maximum ion outflow speed in the jet and the width of the jet
\item Various measurements made by satellites could indicate the relative position of the jet and/or the size of the jet
\end{keypoints}

\begin{abstract}
The structure of dipolarization jets with finite width in the
dawn-dusk direction relevant to magnetic reconnection in the Earth's
magnetotail is explored with particle-in-cell simulations. We carry
out Riemann simulations of the evolution of the jet in the dawn-dusk,
north-south plane to investigate the dependence of the jet structure
on the jet width in the dawn-dusk direction. We find that the magnetic
field and Earth-directed ion flow structure depend on the dawn-dusk
width.  A reversal in the usual Hall magnetic field near the center of
the current sheet on the dusk side of larger jets is observed. For
small widths, the maximum velocity of the Earthward flow is
significantly reduced below the theoretical limit of the upstream
Alfv\'en speed. However, the ion flow speed approaches this limit once
the width exceeds the ion Larmor radius based on the normal magnetic
field, $B_z$.
\end{abstract}

\section{Introduction \label{introduction}}
Magnetic reconnection is a common phenomenon in the Earth's
magnetotail, solar flares, coronal mass ejections, and sawtooth
crashes in tokamaks; as such, it is a subject of active research. When
two oppositely directed field lines reconnect, energy stored in the
magnetic field is transferred to particle energy
\citep{Angelopoulos2008}. The particles are accelerated by the tension
in the reconnected field lines in two oppositely directed jets away
from the x-line, the location where the reconnecting component of the
field vanishes. In the case of the magnetotail, reconnection causes one
jet to move Earthward and the other to be directed
tailward. While it has been shown that the Alfv\'en speed is the
maximum speed to which these dipolarization jets can accelerate flows
\citep{Parker1957}, it is unclear how the extent of the jet in the
dawn-dusk direction affects this result.

Reconnection begins at the x-line and many studies have examined the
ion and electron dissipation regions centered around it.  We focus
farther downstream on the structure of the dipolarization jet.
Previous 2D simulations \citep[e.g.,][]{Sitnov2011,Wu2012} frequently
focus on the $x-z$ plane (in GSM coordinates) and have examined the
snowplow-like behavior of the dipolarization front as it sweeps up
plasma during its earthward motion.  Such simulations implicitly
assume that the dipolarization jet has infinite extent in the
$y$-direction.  Performing fully 3D simulations lifts this
restriction.  Several authors have recently performed such simulations
\citep{Drake2014,Pritchett2014,Sitnov2011,Sitnov2014,Sitnov2017}.
Generally speaking they find that the full dimensionality allows for
more complicated dynamics and the development of structure along the
jet (i.e., in the dawn-dusk direction).  However, the computational
cost associated with such simulations means that it is difficult to
examine the parametric dependence of dipolarization jet evolution.

For example, \citet{Drake2014} studied a fully three-dimensional model
and found that the center of the jet (in the $x$ direction) is fairly
laminar when compared to the Earthward and tailward edges.  This
suggests that simulations in a two-dimensional plane (north-south and
dawn-dusk) could serve as a useful proxy to fully three-dimensional
results.  In the present manuscript we use such simulations to explore
dependence of the structure of magnetotail magnetic reconnection jets
on their finite extent in the cross-tail direction and, in particular,
how this determines the velocity profile of the Earthward-directed ion
flow and the structure of the Hall magnetic fields in the cross-tail
direction.  We therefore are not addressing the physics of
reconnection onset in the magnetotail and the mechanisms that control
the finite extent (in $x$) of these jets.

During reconnection in the magnetotail, ions are accelerated toward the
Earth and down the tail by the newly reconnected magnetic fields. To
accelerate up to the Alfv\'en speed, they first move in the cross-tail
direction, the direction of the reconnection electric field, before they
turn toward or away from the Earth. It is not a surprise, therefore,
that in very narrow jets the ions are unable to gain enough energy to
reach the Alfv\'en speed. In this manuscript we quantify the relation
between the peak ion Earthward flow and the cross-tail width of the
jet.

That the finite cross-tail width of the jet can impact the Hall
magnetic field is more surprising. It is well-known from theory,
simulations and observations that Hall magnetic fields are produced
near the x-line \citep{Oieroset2001,Mandt1994,Sonnerup1979}. These
fields arise because they are primarily frozen-in to the electrons
inside of the ion diffusion region. Thus the electrons, which are
carrying the current in the current sheet, enter the ion diffusion
region and push on the fields, bending the reconnection magnetic field
into the out-of-plane direction. Simulations indicate that these Hall
fields are carried far downstream, $\sim 100 d_i$, from the x-line
\citep{Le2010,Shay2011}, especially away from the center of the
current sheet. We find that the Hall fields reverse inside of the
current sheet for large widths in the
dawn-dusk direction. However, these fields do return to the expected
orientation away from the current sheet. This effect is due to the
ions from the current sheet becoming frozen-in to the magnetic fields
of the jet, and thus pushing on the fields in the opposite direction
of the electrons.

\section{Simulations}\label{sims}
We use the particle-in-cell code {\tt p3d} \citep{Zeiler2002}.  The
magnetic field strength $B_0$ and density $n_0$ define the Alfv\'en
speed $V_{A}=\sqrt{B_0^2/4\pi m_in_0}$, with lengths normalized to the
ion inertial length $d_i =c/\omega_{pi}$, where $\omega_{pi}$ is the
ion plasma frequency, and times to the ion cyclotron time
$\Omega_{i0}^{-1}$.  Electric fields and temperatures are normalized
to $V_{A}B_0/c$ and $m_iV_{A}^2$, respectively. The coordinate system
for all simulations is Geocentric Solar Magnetospheric (GSM).

The initial conditions for the simulations are intended to mimic those
of a dipolarization jet. We start with a classical Harris current
sheet with an asymptotic magnetic field $B_0$, central density $n_0$
and uniform background density $n_b=0.3n_0$. To this equilibrium, we
add a constant $B_z$ that is localized in the dawn-dusk ($y$)
direction with a corresponding plasma density equal to the lobe
density, $n_b$. The intent is to depict late-time tail reconnection in
which the jet is formed from heated lobe plasma. Hence, in the region with
$B_z$ non-zero, the field is non-zero at the symmetry (equatorial)
plane. In order to maintain pressure balance we modify the electron
and ion temperatures in the jet. The temperatures elsewhere, in code
units, are 1/12 and 5/12 for the electrons and ions respectively. The
ion-to-electron mass ratio is set to 25, which is sufficient to
separate the electron and ion scales. All the forces are balanced in the $y$ and $z$ directions but there is an unbalanced force in the $x$
direction due to a non zero $B_z \partial_z B_x$ term in the fluid
momentum equation. This force accelerates the plasma towards the Earth
in the magnetotail. Following initialization the system adjusts and
reaches a quasi-steady configuration as in a conventional Riemann model.

We perform quasi-two-dimensional (quasi-2D) simulations with two full
spatial dimensions ($y$ and $z$) and one ($x$) with small extent.  To mitigate the resulting noise, our analysis of
employs averages over all cells in the $x$ direction. These computationally cheap simulations allow us to explore jets of different widths at late times. Typical simulation dimensions are $(x,y,z)=(3.2,102.4,102.4)d_i$. We also examine a more expensive simulation with $(x,y,z)=(102.4,25.6,25.6)d_i$, which has already been discussed in \citet{Drake2014}. The spatial grid has a resolution $\Delta = 0.05$ while the
smallest physical scale is the Debye length in the magnetotail,
$\approx 0.03$.  We use $100$ particles per cell per species for the
simulations with a small extent in $x$. The simulation with a larger extent in $x$ uses $50$ particles per cell. In all simulations the speed of light is $c=15$ and we employ periodic boundary conditions in all directions. (Note that periodicity in $z$ actually requires us to simulate two mirror-image current sheets. We stop the simulation before the two current sheets begin to interact and only present results from one. Both sheets undergo very similar evolution.) The widths, $w$, we examine are: $\sim 1d_i$, $\sim 2d_i$, $\sim
3d_i$, $\sim 4d_i$, $\sim 7d_i$, $\sim 15d_i$, $\sim 30d_i$, and
infinite (occupying the full space along $y$). For a full comparison between the quasi-2D simulations and the full 3D version see the supporting information.

\section{Presentation and analysis of simulation results}\label{discussion}

\subsection{Reversal of the Hall Field}
\begin{figure}
\centering
\includegraphics[width=27pc]{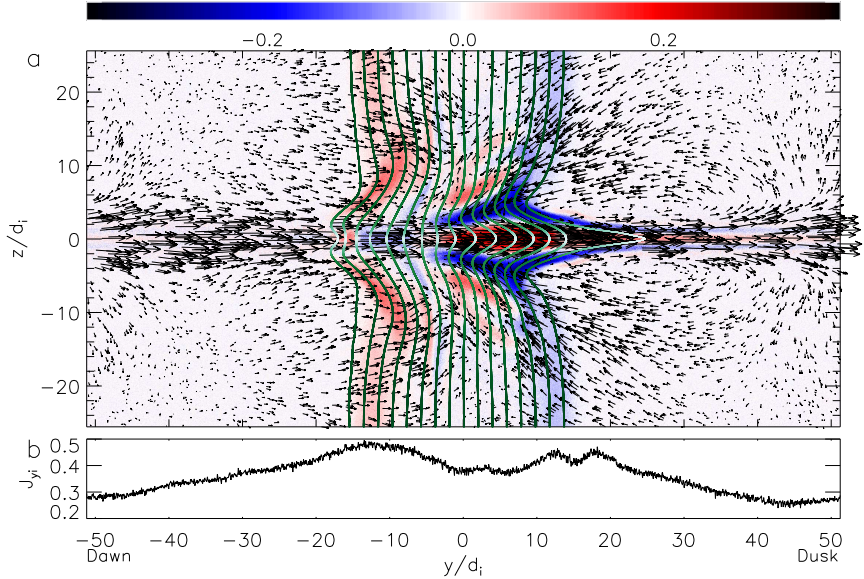}
\caption{Panel (a) is the $x$ component of the total current in the $y-z$ plane at
  $\Omega_{i0}t=40$. Plotted in black arrows are ion flow vectors with the
  length proportional to the magnitude of the velocity.
  The in-plane magnetic field lines in the jet are displayed in green, with darker shades
  indicating a stronger magnetic field. Panel (b) is a cut of the total ion current in the $y$ direction through the center of the current sheet ($z/d_i=0$). }
\label{cs2fieldline_t20}
\end{figure}

As a result of the system's evolution, the Hall magnetic field ($B_y$)
on the dawn edge of the jet reverses sign from its usual direction,
especially for jets with a large width. Figure \ref{cs2fieldline_t20}
is an image of the total current in the $x$ direction for the
simulation with the largest finite-width dipolarization jet, $w \sim
30 d_i$. Overlaid are arrows showing the in-plane ion flows and magnetic field lines in shades of green. The current layer can be 
identified by the ions flowing duskward ($-2 
\leq z/d_i \leq 2$). The electrons (not shown) flow dawnward within the current sheet. The ends of the field lines are
evenly distributed along the bottom of the jet in $z$ and followed
until they leave the area depicted in Figure
\ref{cs2fieldline_t20}. Below the 2D image is a cut through the center
of the current sheet of the ion current in the $y$
direction.

On the dawn (left) side of the jet the electrons carry the magnetic
field to the left and produce the usual sign of the Hall magnetic
field $B_y$ (negative below the center of the current sheet and
positive above). The small dimple on the extreme dawn edge is likely
due to ions carrying the field towards the dusk.

On the dusk (right) side of the jet, however, $B_y$ reverses sign compared with the usual Hall magnetic field, as is visible above and below the center of the current sheet in Figure \ref{cs2fieldline_t20}. The reversal in this case arises due to ions
pushing on the field as a result of the vortex pair that forms just
duskward of the jet. The vortex pair consists of two adjacent vortices
that are centered just above and below the current layer. Their
vorticity has opposite signs so their flows add at the center of the
current layer. The resulting strong duskward ion flow near the center
of the current sheet carries the field lines toward the dusk while the return
(dawnward) ion flow above and below the current sheet carries the
field lines toward the dawn. The result is a reversal in the Hall field
just above and below the center of the current sheet with the 
sign of $B_y$ taking the usual values further away from the current sheet.

We now discuss the mechanism for the development of the vortex
pair. In the initial state the ions flow uniformly across the domain
toward the dusk. The imposed $B_z$ diverts the flowing ions toward the Earth, which slows their duskward flow. The flowing ions
outside of the jet on the dusk side continue to flow toward the dusk and try to fill in this depletion by turning back toward the jet, thereby forming the
two vortices evident in Figure \ref{cs2fieldline_t20}(a). The
reduction of the ion current shown in the cut along the middle of the
current layer in Figure \ref{cs2fieldline_t20}(b) over the interval
$20<y/d_i<40$ results from the diversion of the ion flow away from the
current layer as seen in the velocity vectors of Figure
\ref{cs2fieldline_t20}(a).

The ions in the jet contribute to closing these ion vortices. As time
progresses, the tension in the "reconnected" magnetic field (in the
$z-x$ plane) drives the plasma flow in the positive $x$ (earthward)
direction. This motion causes ions in the lobes to move towards the
current sheet and produces the converging flows (toward $z=0$) seen on
the duskward edge of the jet. Upon entering the current sheet, the
ions move in Speiser-like orbits and rotate first into the positive
$y$ direction before completing their Speiser motion into Earthward
flow. Their motion in the duskward direction as they rotate in the
current layer contributes to the enhancement in $J_{iy}$ seen around
$y/d_i \sim 15$ in Figure \ref{cs2fieldline_t20}(b). This
high-speed ion flow carries the magnetic field duskward
and drives the reversed Hall field seen in the data.

An organized reversal in the Hall magnetic field has not been seen
even in large-scale 2D reconnection simulations (carried out in the
$x$-$z$ plane) \citep{Le2010,Shay2011} although firehose-driven
turbulence in the jet can produce local reversals of $B_y$
\citep{Hietala2015}. An important question then is whether these
earlier results are consistent with the Hall field reversal seen in
the present simulations with the largest jet width. Since nearly all
of the ions in dipolarization jets originates upstream in the lobes,
they carry little cross-tail momentum as they enter the jet and total
momentum conservation in the cross-tail direction means that they will
not gain a significant net flow in the cross-tail direction
\citep{Liu2012}, it is therefore the electrons that carry most of the
cross-tail current in 2D reconnection simulations. An infinite width
jet would, of course, not produce the vortex shown in
Fig.~\ref{cs2fieldline_t20}(a) since $\partial_y=0$. Thus, we would
not expect the Hall field to reverse for jets with infinite extent in
the cross-tail direction. To confirm this, we performed a separate
simulation with an infinite jet with only the electrons carrying the
initial current.  At late times there was a non-zero duskward flow of
ions, but it was significantly smaller than the ion flow depicted in
Figure \ref{cs2fieldline_t20} and was unable to reverse the Hall
field. This result is consistent with past 2D reconnection simulations
that did not see a reversal of the Hall field.

\subsection{The dependence of the Earthward-directed flow on Cross-tail Width}
\begin{figure}
\centering
\includegraphics[width=27pc]{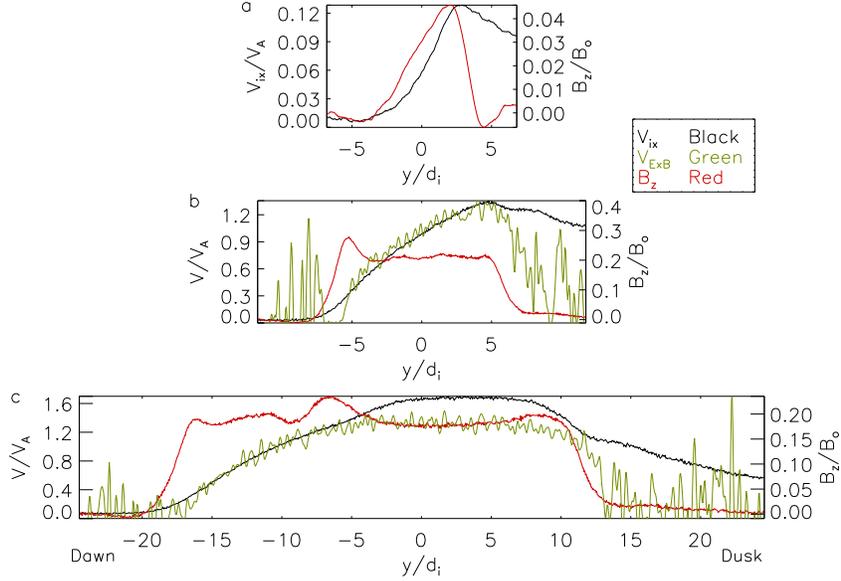}
\caption{The $x$ component of the ion velocity (black) and the $z$
  component of the magnetic field (red) from three runs at
  $\Omega_{i0}t=40$ along a cut taken at $z=0$. (b) and (c) also
  include the local $\mathbf{E} \boldsymbol{\times} \mathbf{B}$
  velocity in the x direction (green). In (a) the smallest jet, width
  $\sim 1 d_i$, in (b) a medium sized jet, width $\sim 15 d_i$, and in (c) the
  largest finite jet, width $\sim 30 d_i$. The images only show a portion
  of the full simulated space. Note that the smallest jet has
  significantly increased in width at late times, but that there is a
  corresponding decrease in the magnetic field to compensate. For this reason the
  $\mathbf{E} \boldsymbol{\times} \mathbf{B}$ velocity is too noisy to
  be seen and thus we have omitted it from this figure.}
\label{VixBz_t21}
\end{figure}

Figure \ref{VixBz_t21} shows the Earthward-directed ion flow velocity
in black through the center of the current sheet ($z/d_i=0$) for jets
with widths of $\sim 1 d_i$, $\sim 15d_i$, and $\sim 30d_i$. For
clarity, the magnetic field profile is also shown in red. For panels
(b) and (c) the local $\mathbf{E} \boldsymbol{\times} \mathbf{B}$
drift velocity is also plotted in green (we neglect this for panel
(a) since it is too noisy). Though not shown, the reconnection
electric field $E_y=V_{E\times B}B_z$ follows the profile of
$V_{E\times B}$ because $B_z$ is nearly constant within the jet. Panel
(c) clearly shows the four phases of the ion outflow along the
cross-tail direction, $y$. First, the ions enter the dawnward side of
the jet with no flow in the $x$ direction. Second, upon entering the
jet, the ions accelerate along the reconnection electric field $E_y$,
which increases with distance into the jet, and rotate into the $x$ or
$\mathbf{E} \boldsymbol{\times} \mathbf{B}$ drift direction. Third,
their Earthward flow plateaus until they exit the jet. In the fourth
phase the ions retain their x-directed motion outside of the jet since
they are moving along the unreconnected magnetic field $B_x$. Outside
of the jet, the ions mix with the background population so that the
outflow speed slowly decays with increasing distance from the
jet. Note that, although only the simulation shown in panel (c) with
$w=30d_i$ is sufficiently large enough to exhibit the plateau, all
three simulations produce an Earthward-directed,
magnetic-field-aligned beam of ions in a region with $B_z=0$.

To understand the scale length of the region where the ions are
accelerating, consider that the ion outflow speed plateaus at the
upstream Alfv\'en speed. Thus we can make the following scaling
argument starting with the steady state ion conservation of momentum
equation for a 2D system ($\partial_x=0$) at $z=0$:
\begin{linenomath*}
\begin{equation}\label{accelerationscalelength}
mnv_{yi}\partial_yv_{xi} \approx \frac{B_z}{4\pi}\partial_zB_x \sim mnv_{yi} \frac{C_A}{L_y} \sim \frac{B_z}{4\pi}\frac{B_x}{\Delta}
\end{equation}
\end{linenomath*}
where $C_A$ is the upstream Alfv\'en speed, $L_y$ is the acceleration length scale, $\Delta$ is the half width of the current sheet, and $v_{zi}=0$ due to the symmetry of the configuration. We neglect the divergence of the pressure since it is smaller than the magnetic tension in the region of interest. Next, since the ions carry most of the current on their entry into the jet, we can approximate $v_y$ from Amp\'ere's law on the dawnside of the jet:
\begin{linenomath*}
\begin{equation}\label{v_y}
\partial_zB_x=\frac{4\pi}{c}nqv_{yi} \sim B_x/\Delta
\end{equation}
\end{linenomath*}
Then, plugging $v_y$ into Equation \ref{accelerationscalelength}, solving for $L_y$ and normalizing to code units we obtain:
\begin{linenomath*}
\begin{equation}\label{L_y}
L_y/d_i \sim \sqrt[]{\frac{n_0}{n_b}}\frac{B_x}{B_z}
\end{equation}
\end{linenomath*}
where $n_b$ is the upstream ion density. For our simulation, $n_b/n_0\approx 0.3$ and $B_x/B_z\approx 5$ giving $L_y/d_i\sim 10$ which is in agreement with Figure \ref{VixBz_t21} panel (c) and explains why panel (b), which only has a width of approximately $10d_i$, has no plateau. 

While the $\mathbf{E} \boldsymbol{\times} \mathbf{B}$ drift describes
the motion of the ions for medium sized and smaller jets (e.g.,
Fig.~\ref{VixBz_t21}(b)), it falls short of accurately depicting the
ion motion for larger jets after they plateau in speed. Thus, it is of
interest to evaluate whether the particle motion within a
dipolarization jet can be described by the usual guiding center
drift.  A quantitative test comes from evaluating the $\kappa$
parameter of \citet{Buchner1989}.  Defined as $\kappa =
\sqrt{R_{\text{min}}/\rho_{\text{max}}}$, where $R_{\text{min}}$ is
the minimum radius of curvature and $\rho_{\text{max}}$ is a
particle's maximum Larmor radius. Values greater than unity are
associated with well-behaved trajectories (i.e., well-described by
drifts). For our simulations we find $\kappa$ for the ions to be
$\lesssim 1$ within the jets. Thus, we cannot
expect the guiding center drifts to accurately describe the observed
flows seen in Figure \ref{VixBz_t21}. Nevertheless, the underlying
physics driving the drifts (curvature of the magnetic field, magnetic
field gradients, etc.)  will still influence the particle flows.

Although there is some agreement between the $\mathbf{E}
\boldsymbol{\times} \mathbf{B}$ drift and the ion flow, there must be additional physics
at work in the region where they diverge (e.g., for $-15\leq y/d_i \leq -5$ in panel
(c)). The curvature of the magnetic field within the jet is a likely
source. Recall from Figure \ref{cs2fieldline_t20} that the Hall field
reverses around $y/di\sim-3$ creating field lines bent duskward. This is also where the ion flow diverges from
the $\mathbf{E} \boldsymbol{\times} \mathbf{B}$ drift indicating that the 
curvature of the magnetic field boosts the ion flow in the $x$ direction. 
Direct calculation of the curvature drift leads to ion flow velocities
significantly larger than observed (although of the correct sign).
However, because $\kappa \lesssim 1$, the particle orbits are not
well-described by drifts superimposed on circular Larmor orbits and
agreement should not be expected.

It should be noted that although the jet in panel (b) has a width
$\sim 2$ times greater than that of panel (a) at the time shown, the
jet widths at $t=0$ differed by a factor of $\sim 15$. The widening of
the jet in panel (a) is accompanied by a drop in $B_z$, so that the
integrated magnetic flux is constant.  Once the simulation begins, the
electrons, as the less massive species, interact with the fields
first. The electrons in the region inside of the jet initially have a
large velocity in the dawn direction and a low density compared to the
electrons in the current sheet outside of the jet.  Hence the
electrons in the jet carry the magnetic flux $B_z$ in the jet toward
the dawn, while the electrons entering the jet from the dusk carry the
flux with a lower velocity. The resulting divergence of the cross-tail
electron flow spreads the flux as seen in Fig.~\ref{VixBz_t21}(a) and
leads to a build-up of magnetic flux on the dawn (left) side for
larger jets. This effect leads to the smallest initial-width jet
achieving a width comparable to the medium sized jet (panels a and b
in Figure \ref{VixBz_t21}).

\subsection{Maximum Ion Flow Dependence on Jet Width}
It is well known that the ion flow in a dipolarization jet can
be accelerated to the upstream Alfv\'en speed in a jet that has
infinite cross-tail width. For the case of a finite width, jet we have
shown that there is a transition between the nearly zero ion flow
outside of the jet on the dawn side and where the flow reaches the
maximum outflow speed. An ion in the current sheet enters a region of
non-zero $B_z$ and $E_y$ from the dawn side and gains an $\mathbf{E}
\boldsymbol{\times} \mathbf{B}$ drift in the $x$
direction. It is reasonable to expect that jets with a width
less than the ion Larmor radius in the magnetic field, $B_z$, would be
unable to fully accelerate the ions to this maximum speed since they
would leave the jet before completing a full orbit. In our simulations
the ions have a local Larmor radius of $\sim5 d_i$ (due to $B_z$) in
the center of the jet. We start from Equation \ref{accelerationscalelength}. Since the jet widths are short, the ion velocity $v_{yi}$ inside the jet is nearly unchanged from that outside so $v_{yi}$ is again given in Equation \ref{v_y}. The resulting equation for $v_{xi}$ is as follows: 
\begin{linenomath*}
\begin{equation}
mn\left(\frac{c\partial_z B_x}{4\pi nq}\right)\partial_yv_{xi}\approx \frac{B_z}{4\pi}\partial_zB_x \sim mn\left(\frac{cB_x}{4\pi nq \Delta}\right)\frac{\partial v_{xi}}{\partial y} \sim \frac{B_zB_x}{4\pi \Delta},
\end{equation}
\end{linenomath*}
where $\Delta$ is the half width of the current sheet. The equation can be integrated to obtain the increase in $v_{xi}$ with distance inside of the jet. However, the integral is complicated by the fact that $B_z$ spreads in time as shown in Fig.~\ref{VixBz_t21}. Fortunately, the total magnetic flux associated with $B_z$ is a constant and is given by $B_{z0}w$, where $B_{z0}$ is the initial value of $B_z$. Completing the integral, we obtain the maximum ion flow speed inside the dipolarization jet $v_{xi,max}$. In normalized code units the result is given by
\begin{linenomath*}
\begin{equation}
v_{xi,max}/V_A \sim \frac{B_{z0}}{B_x} w/d_i
\end{equation}
\end{linenomath*}

\begin{figure}
\centering
\includegraphics[width=27pc]{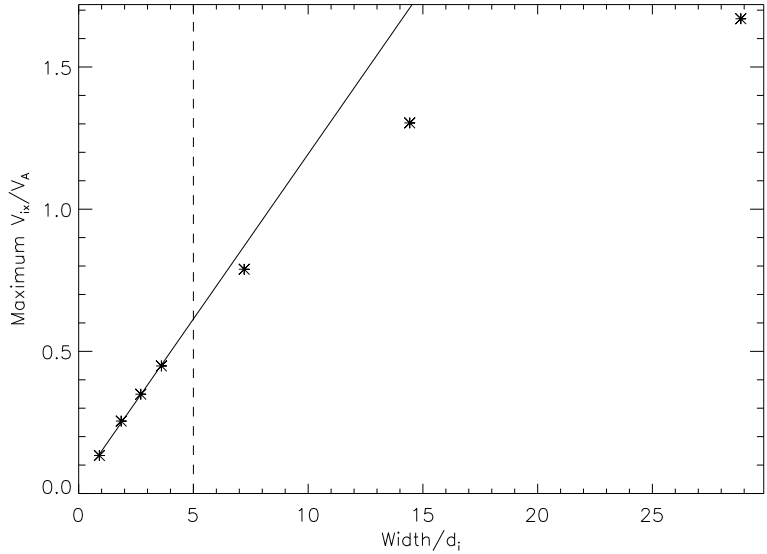}
\caption{The $x$ component of the maximum ion flow velocity inside the
  jet versus the width. The linear fit shown in this figure (solid) was
  calculated from the first four data points. Widths greater than the local Larmor radius (dashed), $\sim 5 d_i$, are no longer linear.}
\label{linvi}
\end{figure}

Taking values from the simulations for $B_z(=0.2)$, and $B_x(=1)$, we
arrive at a slope of 0.2 for $v_{xi,max}/V_A$ versus $w/d_i$.
Figure \ref{linvi} shows the relationship between the maximum outflow
velocity and the width $w$. A linear fit to the first few data
points gives a slope of $\sim0.12$ which differs by less than a factor
of $2$ from our simple model. At larger widths, the jet velocity approaches
the upstream Alfv\'en speed with increasing $w$.
The simulation results deviate from the linear relationship for the
simulation with $w=7d_i$, suggesting that other physical effects
become important around this length scale.  This is expected since the
data from $w=4d_i$ and $w=7d_i$ bracket the local ion Larmor radius of
$5d_i$ in the jet (i.e. bracket the length scale at which we expect
our model to break down). For $w$ well above $\sim 5 d_i$ the maximum
$v_{ix}$ then approaches the upstream Alfv\'en speed, $\sim 1.8
V_A$. Figure \ref{linvi} clearly shows the strong dependence of the
speed of the flow on the width of the jet.

\section{Conclusion}\label{Conclusion}
We use quasi-2D particle-in-cell simulations to analyze the structure
of the ion outflow and magnetic field on the dawn-dusk width of a dipolarization jet. For large, finite widths there is a significant
reversal in the Hall magnetic field on the dusk side of the jet. This
reversal, which should be detectable by spacecraft observations,
results from the interaction of duskward moving ions with the magnetic
field and results in the generation of a $y$ component of the
magnetic field. The reversal happens approximately where the
$x$-component of the ion flow reaches a maximum inside the jet and is
supported by vortices above and below the current sheet that extend
outside of the jet on the dusk side. The reversal of the Hall magnetic
field does not appear for jets that have infinite extent in the
cross-tail direction, which is consistent with simulations of
symmetric reconnection that reveal Hall magnetic fields of the normal
sign far downstream from the x-line \citep{Le2010,Shay2011} and
simulations that do not show any coherent behavior near the current
sheet \citep{Hietala2015}. The bent field lines shown in our
simulations are due to the ions moving with high velocity in the dusk
direction bending the magnetic field $B_z$ into the cross-tail
direction.  This behavior does not appear in 2D models of reconnection
because the ions initially in the current sheet and which carry large
momentum in the cross-tail direction are swept downstream in a
plasmoid \citep{Arzner2001} leaving only a population of ions from the
lobes inside the jet.  These ions are unable to interact with the
current sheet and form a flow with a high enough cross-tail speed to
reverse the Hall field.

In addition, we were able to demonstrate that there is a strong
dependence of the maximum Earthward flow on the width of a dipolarization jet
in the dawn-dusk direction. For jets smaller than the local Larmor
radius in the reconnected magnetic field $B_z$ there is a linear
relationship between the width of the jet and the maximum ion outflow
velocity inside the jet. At larger widths, the maximum ion outflow
velocity then approaches the upstream Alfv\'en speed. A scaling
argument in support of this linear relationship suggests that the
slope only depends on the ratio of $B_z$ to the upstream $B_x$.

The results of this paper could be used to establish the relative
positions of satellites with respect to a dipolarization jet as well
as placing limits on the cross-tail width of the jet. If the ion flow
in the $x$ direction is faster than the local $\mathbf{E}
\boldsymbol{\times} \mathbf{B}$ drift, especially in the presence of a
reversal in the usual Hall field, the satellite must be on the dusk
side of the jet and the width must be larger than the
scale length $L_y/d_i=\sqrt{n_0/n_u}(B_x/B_z)$. Alternatively,
if an ion flow in the $x$ direction is observed without a finite
$B_z$, that would indicate a position duskward of a dipolarization
jet. A measured jet outflow velocity well below the upstream Alfv\'en
speed and with $B_z\neq 0$ would suggest that the satellite is on the
dawnward side of the jet or that the jet has a very small cross-tail
width compared with the effective scale length
$L_y/d_i=\sqrt{n_0/n_u}(B_x/B_z)$.

\begin{acknowledgments}
 This work was supported by NASA grant NNX14AC78G. The simulations were carried
 out at the National Energy Research Scientific Computing Center. The
 data used to perform the analysis and construct the figures for this
 paper are available upon request.

\end{acknowledgments}

\end{document}